\documentstyle[prd,preprint,aps]{revtex}

\newcommand{\be}{\begin{equation}}
\newcommand{\ee}{\end{equation}}
\newcommand{\bea}{\begin{eqnarray}}
\newcommand{\eea}{\end{eqnarray}}

\newcommand{\A} {{\cal A}}

\newcommand{\EP} {E_0^{\prime}}
\newcommand{\EPP} {E_0^{\prime\prime}}

\newcommand{\LL} {{\cal L}}

\newcommand{\EDD} {{\ddot E}_0}
\newcommand{\EDP} {{\dot E}_0^{\prime}}

\newcommand{\BBB} {\left(n+\alpha m^p\right)}
\newcommand{\QQ} {\Psi_{0}}

\begin{document}

\reversemarginpar
\tighten

\title{Questions of Stability near Black Hole Critical Points}

\author{Gilad Gour\thanks{E-mail:~gilgour@phys.ualberta.ca}
and A.J.M. Medved\thanks{E-mail:~amedved@phys.ualberta.ca}}

\address{
Department of Physics and Theoretical Physics Institute\\
University of Alberta\\
Edmonton, Canada T6G-2J1\\}

\maketitle

\begin{abstract}

In this letter, we discuss how thermal fluctuations can
effect  the stability of (generally) charged  black holes 
when close to certain  critical points.
Our novel treatment utilizes 
 the black hole area spectrum
(which is, for definiteness, taken to be evenly spaced)
 and makes an important distinction between   fixed and fluctuating
charge systems (with these being modeled, respectively, as a canonical
and grand canonical ensemble.)
The discussion begins with a summary of 
a recent  technical paper \cite{NEW}.  We  then go on to
consider the issue of stability when the system approaches
the critical  points of interest. These include
 the $d$-dimensional analogue of the Hawking-Page phase transition,
a phase transition that is relevant to Reissner-Nordstrom
black holes
and various extremal-limiting cases.
\\
\end{abstract}

It has often been suggested that any viable theory
of quantum gravity must somehow account
for black hole thermodynamics \cite{smo}. In particular, a prospective
fundamental theory would be expected to
provide a statistical explanation for
the Bekenstein-Hawking  entropy \cite{bek1,haw1}:
$S_{BH}={1\over 4G}A$, where $A$ is the
surface area of the black hole horizon.\footnote{Here and throughout,
all  fundamental constants - besides the gravitational coupling constant, 
$G$ - are set to unity.}

An important subplot
is the leading-order quantum correction
to this classical area law.
Indeed, there has been substantial interest
in deducing this  correction, with many
different techniques having been called upon for
just this purpose (see \cite{NEW} for 
a thorough list of references).
A  feature that is common to all of the relevant studies
is a leading-order correction
that arises at the logarithmic order in $S_{BH}$.
Nonetheless, there is significant disagreement
on the value of the proportionality constant
or logarithmic {\it prefactor}.

This conflict over the prefactor can be partially resolved
 when one considers an often overlooked point:
 there  are actually  two distinct sources
for this logarithmic correction.
More precisely, one  can anticipate both an   
uncertainty  
in the number of microstates  describing  a black hole
with a fixed geometry {\it and}
a quantum correction due to  thermal fluctuations
in the area of the horizon.  
In principle, these two contributions should be completely separable,
at least to the logarithmic order \cite{can,gg2}.
For the remainder, we will  focus our
attention on the latter (thermal) correction.

It has sometimes been implied that the thermal correction
can be obtained, given a suitable canonical framework, 
without any reference to the fundamental
theory of quantum gravity. (This clearly can not be
the case for the other, microcanonical correction, which ultimately
 must  refer  to the underlying  degrees of freedom.)
However, in a recent paper \cite{NEW}, we have argued
that an important element of the quantum theory  -
namely, the black hole area spectrum -  must
inevitably be accounted for 
in any such calculation. With this realization in mind, we
proceeded  to calculate the thermal correction to the black hole entropy; both
generically and for an assortment of special
models (in particular, various limiting cases
of a Reissner-Nordstrom black hole in an 
anti-de Sitter, or AdS, spacetime of arbitrary dimensionality).
 
For the sake of definiteness, we made the decision in  \cite{NEW}
to work with an evenly  spaced area spectrum (and will continue
to do so here). Although this spectral form  remains somewhat
controversial, let us take note of the 
support in the literature; beginning with 
the heuristic arguments of Bekenstein \cite{bek2}
and more rigorous treatments since  
({\it e.g.}, 
\cite{bek3,mak1,kun1,bek4,gj1}).\footnote{It 
is also noteworthy that, even in the context of
loop quantum gravity, certain
authors have advocated for an equally spaced spectrum  \cite{lqg}.}

Another novelty of our prior work \cite{NEW} is that
a crucial distinction was made between
black holes with a fixed charge
and those for which the charge
is allowed to fluctuate. (The charge can
legitimately  be regarded as a fixed quantity  only under
certain circumstances; for instance, a black hole
 confined within  a neutral heat bath.)
 Accordingly, we modeled the  system as both
a canonical ensemble  and a
{\it grand} canonical ensemble (respectively). 
The formal differences  will be further  clarified
below.

The remainder of the letter is organized as follows.
We  first  give a summarized account of
the preceding work \cite{NEW}, as this
establishes the foundation for  the subsequent analysis.
Next, we investigate the  issue of black hole stability 
when the system approaches certain critical points (discussed
thoroughly below). The letter ends with a brief conclusion. 

In review of  \cite{NEW},  let us first consider the case of a black
hole with a fixed charge. Given the premise of
a black hole in a state of thermal equilibrium 
({\it i.e.}, a black hole in a ``box''),  an appropriate
model for the system is a canonical ensemble
of particles and fields. Under the assumption of 
an evenly spaced area spectrum, $A(n)\sim n$ ($n=0,1,2,...$),
 and a semi-classical (or large black
hole) regime, it follows that the  canonical
partition function  can be expressed as
\be
{\cal Z}_C(\beta)
=\int^{\infty}_{0} dn \exp\left(-\beta E(n)+\epsilon n\right)\;.
\label{121}
\ee
Here, $\beta^{-1}$ is the fixed temperature and
 $E(n)$ is the energy of the $n$-th level. Furthermore, the degeneracy  
 has been fixed by way of the
entropic area law; that is,  $S(n)=\epsilon n$, with $\epsilon$ being
a dimensionless, positive parameter of the order unity.

Evaluating the above and then employing text-book
thermodynamics, we found the following
for the canonical entropy:
\be
S_C\approx S_{BH} 
-{1\over 2}\ln\left[{\EPP\over \EP}\right]\;,
\label{129}
\ee
where a prime (always) denotes differentiation with respect
to $n$, a subscript of $0$ indicates a quantity
evaluated at $n=n_0\equiv <n>$, and  
 the classical black hole entropy
has been identified as $S_{BH}=\epsilon n_0$.
(An approximation sign always signifies that
irrelevant constant terms and higher-order corrections 
have been neglected.) 

Take note of the right-most term, which is
the anticipated logarithmic correction to
the classical area law.
As an immediate consequence of the formalism,
the argument of this logarithm 
must be strictly positive. Given that
the temperature can not be negative (by
virtue of cosmic censorship) and that $\EP=\epsilon\beta^{-1}$
(via the first law of  thermodynamics),
one is  able to identify a pair of {\it stability}
conditions:  $\EP>0$ and   $\EPP>0$.

The canonical partition function  can
also be used to evaluate the variation
of the spectral number. For this calculation, we have obtained 
\be
(\Delta n)^2 \equiv <n^2>-<n>^2 \approx
{\EP\over\epsilon\EPP}\;.
\label{6}
\ee
Keep in mind that this variation serves as a
direct measure of  the thermal
fluctuations in the area  or entropy;
that is, $\Delta S_{BH}\sim \Delta A\sim \Delta n$.
Consequently, Eq.(\ref{129}) can be
elegantly rewritten as
\be
S_{C}\approx S_{BH}+\ln[\Delta S_{BH}]\;.
\label{entflu}
\ee

Secondly, let us consider the case of
a black hole with a fluctuating charge.
There should now be an additional  quantum number
 that accounts for  this dynamical charge 
and it is, therefore,  appropriate to model the system
as a grand canonical ensemble. 
 Given a uniformly  spaced area
spectrum and the independence of the quantum numbers, 
we were able to deduce the following spectral form:
\be
A(n,m)\sim n+\alpha m^p\;,\quad\quad\quad 
 n,|m|=0,1,2,...\;,
\label{blah1}
\ee
where the ``new'' quantum number, $m$, directly measures
the black hole charge according to $Q=me$ (with $e$ being a fundamental
unit of electrostatic charge). Also, 
$\alpha$ is a positive constant and $p$ is a positive, rational  number.

We are now in a position to write down the 
grand canonical partition function,
\be
{\cal Z}_G(\beta,\lambda)=\int_{-\infty}^{\infty}dm\int_{0}^{\infty}
dn 
\exp\left(-\beta E(n,m) +\epsilon\left(n+\alpha m^p\right) +\lambda m\right)\;,
\label{192}
\ee
where $\lambda$ is a chemical (or electric) potential
and $\epsilon$ is (again) a dimensionless, positive parameter.

After some lengthy evaluation, the {\it grand}  canonical entropy
was eventually found to be as follows:
\be
S_G \approx S_{BH} 
-{1\over 2}\ln
\left[{\QQ\over (\EP)^2}
\right]\;,
\label{13}
\ee
where we have defined
\be
\QQ\equiv \EPP\left(\EDD-\alpha p(p-1)m_{0}^{p-2} \EP\right)
-(\EDP)^2\;
\label{blah2}
\ee
and identified
\be
S_{BH}={\epsilon}\langle n+\alpha m^p\rangle.
\label{blah3}
\ee
Furthermore, a subscript of 0 now indicates
a quantity evaluated at {\it both}  $n=n_0$ and $m=m_0\equiv <m>$,
and a dot represents a differentiation with respect to $m$.

In view of the revised form of the logarithmic correction (\ref{13}),
the relevant stability conditions  now become  $\EP>0$ and 
 $\QQ>0$. (Note that
the relation $\EP=\epsilon\beta^{-1}$ is still valid.)

As for the prior (canonical)  treatment, the grand canonical partition
function can be used to evaluate the variations in the quantum numbers:
\be
(\Delta n)^2 \approx {\EP\left[\QQ+(\EDP)^2\right]\over 
\epsilon\EPP\QQ}\;,
\label{19}
\ee
\be
(\Delta m)^2 \approx 
 {\EP\EPP
\over \epsilon\QQ}\;.
\label{232}
\ee

In the work of interest \cite{NEW},
we also translated our generic formalism into the framework of a
 $d$-dimensional ($d\geq 4$) AdS-Reissner-Nordstrom
black hole. Beginning with  the defining relation for the horizon
\cite{cejm} and the first law of black hole mechanics, we could
readily obtain the spectral form of the energy. 
In this regard, it is useful to note
that $R(n,m)\sim [A(n,m)]^{1\over d-2}$  where
$R$ is the radius of the horizon and $m$ may or may not
be fixed.

For the case of a fixed charge, one finds that 
\be
E(n)= {1\over \omega_d}\left[{\LL^{d-1}\over L^2}n^{{d-1\over d-2}}
+\LL^{d-3} n^{{d-3\over d-2}} +{\omega_d^2 Q^2\over\LL^{d-3}}
n^{-{d-3\over d-2}}
\right]\;,
\label{160}
\ee
where $L$ is the  AdS curvature parameter, 
$\omega_d$ is a scaled form of the $d$-dimensional
Newton constant, and $\LL$ is a length scale 
such that $R_0^{d-2}=\LL^{d-2} S_{BH}/\epsilon$.
It should be kept in mind that  the 
curvature parameter, $L$,
can always be viewed as  the effective box size
({\it i.e.}, the spatial extent) of the system.

Meanwhile,
for the fluctuating-charge scenario,
the spectral energy can be expressed  as follows:
 \be
E(n,m)  =  
\frac{1}{\omega_d}\left[\frac{\LL^{d-1}}{L^2}\A^{\frac{d-1}{d-2}}
+\LL^{d-3} \A^{\frac{d-3}{d-2}} +\frac{\omega_d^2 e^2 m^2}{\LL^{d-3}}
\A^{-\frac{d-3}{d-2}}
\right] \;,
\label{239}
\ee
where $\A\equiv \BBB$ is  the dimensionless area.
Take note of the implicit appearance of the parameter $p$ in the above
expression. For some specific cases, it is possible
to fix $p$  by an inspection of the classical black hole
area, as a function of charge, at extremality.
(In particular, $p=1$ if  $R>>L$ and $p={d-2\over d-3}$
if $R<<L$ \cite{NEW}.)
It is, however, an interesting feature of this
Reissner-Nordstrom model that most of the results 
do not depend {\it explicitly} on  $p$. This oddity follows from  
a  calculation of $\QQ$ (\ref{blah2}), which
shows that, if  we write $E(\A)=f(\A)+m^2g(\A)$, then
$\QQ =2g_0f_0^{\prime\prime}
+2m_0^{2}g_0g_0^{\prime\prime}-4m_0^{2}(g_0^{\prime})^{2}$.

Given the above expressions, it becomes a straightforward
process to determine the thermal correction to the entropy
(as well as the thermal fluctuations) for any 
AdS-Reissner-Nordstrom  black hole.  In the prior work,
we considered a number of interesting limits, where
the calculations are most easily interpreted.
Before summarizing these findings, let us
point out that, in all cases considered,
the canonical (grand canonical) entropy takes
on a particularly  simple form,
\be
S_{C(G)}=S_{BH}+b\ln[S_{BH}]\;,
\label{blah4}
\ee
where $b$ is a non-negative, rational number that
is independent of the dimensionality of the spacetime.
Except  for the independence of $d$, we expect
that this form  persists under more general
circumstances. 

Let us now recall, case by case, some of the more
prominent outcomes of \cite{NEW}.

{\bf (i) $L<<R$ and $Q\sim 0$}: This case can be viewed
as a neutral black hole in a small box. When the 
charge is fixed, the logarithmic prefactor is
$b=1/2$;\footnote{As a point of interest, the exact same
value, $b=1/2$, was found for the (3-dimensional) BTZ black
hole model \cite{btz}, irrespective of the curvature parameter.}
 whereas $b=1$ if the charge is
allowed to fluctuate. This implies that each
quantum number - that is,  each freely fluctuating parameter -
induces a thermal correction of precisely 
${1\over 2}\ln S_{BH}$. (Let us emphasize that, for
a fluctuating-charge scenario, the condition $Q\sim 0$
should really be written as $<Q>\sim 0$.  
Regardless of this neutrality, the  fluctuations
can still be,  and generally are, quite large.)
As for the variations in the quantum numbers,
we found that, for a fixed charge, $\Delta n\sim n^{1\over 2}_0$ and, when
the charge is fluctuating, $\Delta n ,\Delta m \sim \A_0^{1\over 2}$.
Since $p=1$ in this limit (see above),
it follows that, for both scenarios, $\Delta S_{BH}\sim S_{BH}^{1\over 2}$.

{\bf (ii) $L\sim R$ and $Q\sim 0$}:
It turns out that, for a sufficiently large $L$  and vanishing charge, the 
stability condition $\EPP>0$ (or the analogue, $\QQ >0$) will be violated.
The maximal value of box size (defined by the saturation
of the stability condition),
\be
L_{max}\equiv 
\sqrt{d-1\over d-3}R_0 =
\sqrt{d-1\over d-3}\LL n_0^{1\over d-2}\;,
\label{152}
\ee
can be   identified as
the $d$-dimensional analogue of the Hawking-Page 
phase-transition point \cite{haw4}.
By way of a perturbative expansion, we have shown that,
near the transition point, $b=1$.  Also, $\Delta n\sim n_0$
when the charge is fixed and $\Delta n\sim \A_0$
for the dynamical-charge scenario. Moreover,
it can be readily verified that, near this point,
 the charge fluctuations are always suppressed:
 $\Delta m\sim$ constant. However, the rather large
fluctuations in the quantum number $n$ (and, hence, in
the area) are indicative of an instability in the system. This
notion will be elaborated on  below.

{\bf (iii) $L<<R$ and $Q\sim Q_{ext}$}:
Here, we are using $Q_{ext}$ to denote
the charge of an extremal black hole. Significantly,
 the extremal limit is also the limit of vanishing temperature
({\it i.e.}, $\beta^{-1}=0$)
and, therefore, signifies the saturation of the stability
condition $\EP>0$. Hence, $|Q_{ext}|$ serves
as an upper bound on the magnitude of the charge.
Using a perturbative expansion, we found that,
close to extremality, the logarithmic prefactor
vanishes ($b=0$) and all fluctuations
are completely suppressed ($\Delta n,\Delta m\sim$ constant).
This suppression can be viewed as a natural mechanism
for enforcing cosmic censorship in a near-extremal
regime.

{\bf (iv) $L>>R$ and $Q\sim Q_{ext}$ and} fixed charge: 
First note that, in this regime of large box size
(or, equivalently, the asymptotically flat space limit),
the fixed and fluctuating charge scenarios
turn out to be  qualitatively much different, and so these
will be dealt with as  separate cases.
In the current (fixed-charge) case, the thermal fluctuations
and logarithmic prefactor
were shown, once again, to be completely 
suppressed when the black hole is near extremality.

{\bf (v) $L>>R$ and $Q\sim Q_{min}$ and} fixed charge:
By virtue of the stability condition $\EPP>0$,
it can be demonstrated that (when $L$ is large) there is also a 
{\it lower} bound
on the charge. 
Quantitatively,
one finds that
\be
Q^2_{min}\equiv 
{R^{2(d-3)}\over (2d-5)\omega^2_d}
= {\LL^{2(d-3)}\over (2d-5) \omega^2_d} n_0^{2(d-3)\over d-2}\;,
\label{173}
\ee
and note that 
$Q^2_{min}={1\over 2d-5}Q^2_{ext}\;$.\footnote{Although $Q^2_{ext}$
and $Q^2_{min}$ are relatively  close in units of black hole charge,
it should be remembered that, for a semiclassical
black hole, $Q^2_{ext}$ is a very large quantity
in Planck units, and so cases {\it iv} and {\it v} can  safely be viewed
as isolated regimes.}  This point of minimal
charge, but finite temperature, is certainly indicative of
a phase transition. (The existence of
such a  phase transition for  Reissner-Nordstrom
black holes was  first  established by  Davies \cite{dav}.)
In fact, close to the minimal charge, the ensemble behaves
in precisely the same way as it does near
 the Hawking-Page  transition point ({\it cf}, case {\it ii}).
That is, the logarithmic prefactor is quite large, $b=1$,
and the area fluctuates according to  $\Delta n\sim n_0$.
Again, such behavior is suggestive of an instability
in the system near the critical point.

{\bf (vi) $L>>R$ and} dynamical  charge:
For this case, we have shown that it is impossible
to simultaneously satisfy both of the relevant stability
conditions, $\EP>0$ and $\QQ>0$.  In the language
of the prior case, this instability is  a consequence of the
coincidence $Q^2_{ext}=Q^2_{min}$. 
Hence, when the charge is allowed to fluctuate,  a non-extremal black hole 
in a large box can {\it not} possibly be stable.

We would now like to extend the prior analysis by
asking (and then answering) the following questions:
just how close can the system approach the  extremal  limit
({\it cf}, cases {\it iii} and {\it iv}) and the phase-transition 
points ({\it cf}, cases {\it ii} and {\it v}) before instability
ensues? 
With regard to the former, given the  suppression of the thermal fluctuations,
one might naively expect that
a near-extremal black hole can come infinitesimally close
to its point of extremality. (Stability near the phase-transition points
is somewhat  less clear, inasmuch  as these thermal fluctuations
are far from suppressed.)
However, we will argue below that this, in fact,
 can {\it not} be the case.\footnote{Note
that our formalism does break down 
at, precisely, any of  the critical points under consideration. Hence, we
can not comment on  the nature
of a {\it perfectly} extremal black hole, assuming that such
an entity can even exist. For  discussion and
references on this controversial issue, see \cite{kun2x}.}

To help answer these questions, let us work
under the following sensible premise:  stability necessitates
that the thermal fluctuations can {\it not} be any larger
than the {\it deviation} from the relevant critical point.  
We will show how this notion can be put in more rigorous
terms upon  examining some specific cases.

Let us proceed by first considering case {\it ii} for a
(neutral) black hole near the Hawking-Page
phase-transition point. 
For simplicity, we will, for the time being, focus on the fixed-charge
scenario.
Recall that the phase transition takes place when $L$
reaches a maximal  value.
 It is, however, more appropriate to
keep  $L$ as  a  fixed parameter and  consider
changes in the dynamical parameter $n$.
 In this spirit,  Eq.(\ref{152}) can be re-expressed as
a minimal (or critical) bound, $n_c$, on the expectation value of $n$.
That is,
\be
n_0 > n_{c}\equiv\left[\sqrt{d-3\over d-1} {L\over\LL}\right]^{d-2}\;.
\label{blah} 
\ee
Since  the system is supposed to be in the ``proximity''
of this transition point, it is appropriate
to write
\be
n_0=n_c+\varepsilon\;, \quad\quad\quad {\rm where}
\quad\quad\quad \varepsilon\equiv n_c^{\gamma}\delta\;,
\label{blah5}
\ee
such that $\gamma<1$ and $\delta$
is a constant, positive parameter of the order unity.
Note that $\varepsilon$ represents the ``deviation''
that was alluded to above.

Given the above expansion (and keeping in mind that $Q\sim 0$),
some straightforward calculation shows  that, to lowest order
in $\varepsilon$,
$\EP\sim n_c^{-{1\over d-2}}$,
$\EPP\sim  n_c^{-{2d-3\over d-2}}\varepsilon$, and so
\be
\Delta n \approx \sqrt{\EP\over \EPP}\sim {n_c\over\sqrt{\varepsilon}}\;.
\label{blah6}
\ee

Now let us reconsider our {\it ansatz} for stability.
In quantitative terms, this essentially means that $\Delta n< \varepsilon$,
which translates into $n_c<\varepsilon^{3\over2}$ or, equivalently,
$\gamma>2/3$ ({\it cf}, Eq.(\ref{blah5})).
For a macroscopically large black hole ({\it i.e.}, $n_0$ or $n_c>>1$),
$n_c^{2/ 3} << n_c$, so that the deviation ($\varepsilon$)
can  be regarded as  small relative to the size of the black hole.
Nonetheless, the minimum  deviation is still a very large
number in Planck units and establishes our previous claim
regarding instability near the Hawking-Page phase-transition point.
As a further point of interest, one obtains
$b=2/3$ for the logarithmic prefactor 
when the system is  ``teetering'' on instability.

One can apply the same general technique to
the fluctuating-charge version of case {\it ii} and obtain very similar
results. (It is not necessary to fix the parameter $p$, so
long as $p\geq1$. Also note that, for completeness, one should
  compare
both $\Delta n$ and $\Delta m$ with $\varepsilon$.)
It is interesting to note that $\Delta m\sim\varepsilon$,
 so that  both   $\Delta n$ and   $\Delta m$  go as
$n_c^{2/ 3}$ near the point where instability ensues. 

One can also apply this methodology to the 
Reissner-Nordstrom phase transition of case {\it v}. To recall,
the stability condition $\EPP>0$ necessitates
that, for a black hole in a large box, 
 there is a minimum value of {\it fixed} charge and, hence,
a maximum value of $n_0$ ({\it cf}, Eq.(\ref{173})). 
In this case, the same approach as above leads to
{\it precisely} the same constraint on the deviation ($\varepsilon$); that is,
$\gamma >2/3$, where $\varepsilon \sim n_c^{\gamma}$
and  $n_c$ is now the critical point relevant to 
case {\it v}.
To the best of our knowledge, this is the first
mention of such a similarity between two
otherwise unrelated phase transitions \cite{haw4,dav}.

Let us next ponder the question of stability 
for  near-extremal black holes. For
illustrative purposes, we will specifically concentrate
on case {\it iv} (large $L$  and  fixed $Q$); nonetheless, 
the same features can be shown to persist
for {\it any} of the near-extremal models of
interest.\footnote{As one might imagine, there
are additional complications that arise when
the charge is fluctuating; {\it cf}, case {\it iii}.
Here, it is useful to fix  $p=1$,
as is appropriate for a small $L$ regime.
Then it follows that, close to extremality, $\A_0\sim m_0$,
so that both $\A_0$ and $m_0$ can be expanded
in terms of $|m_0|=|m_c|-\varepsilon$, where
$m_c$ is the critical (extremal) value.
Also note that, for the purpose of assessing stability, 
both $\Delta n$ and $\Delta m$
need to be compared with $\varepsilon$.}
  
Given that $L>>R$ (and so 
the first term in Eq.(\ref{160}) can safely be
disregarded), it is not difficult to solve
for the extremal value of charge and
then re-express this as a critical minimum, $n_c$, on
the expectation value of $n$. Following this procedure,
we obtain
\be
n_0 > n_{c}\equiv\left[\omega_d |Q|\over\LL^{d-3}\right]^{{d-2\over d-3}}\;.
\label{blah7} 
\ee
As before, we can  appropriately write
\be
n_0=n_c+\varepsilon\;, \quad\quad\quad {\rm where}
\quad\quad\quad \varepsilon\equiv n_c^{\gamma}\delta\;,
\label{blah8}
\ee
with the understanding that $\gamma<1$ and $\delta\sim{\cal O}[1]$.

Utilizing  the above expansion (and remembering  that $L>>R$),
we can deduce, to lowest order
in $\varepsilon$,
$\EP\sim n_c^{-{d-1\over d-2}}$,
$\EPP\sim  n_c^{-{d-1\over d-2}}\varepsilon$, so that
\be
\Delta n \approx \sqrt{\EP\over \EPP}\sim \sqrt{\varepsilon}\;.
\label{blahx}
\ee

It is clear that the thermal fluctuation, $\Delta n$, goes to zero
as $\varepsilon\rightarrow 0$. Hence, we can
now set $\gamma =0$ without endangering stability,
and so $\varepsilon=\delta$. The pertinent question
becomes is there any lower limit on $\delta$? In fact, by imposing
our stability constraint, $\Delta n< \varepsilon$, we
must have $\delta >1$. This means that there
is, indeed,  a natural limit on just how close a near-extremal black hole
can approach  absolute extremality; roughly (but not less than)
one Planck unit of area.  It is reasonable to suggest that
this censoring mechanism is a  manifestation
of some  {\it generalized} form of the  third law of thermodynamics.
Let us re-emphasize that the same censorship
occurs for  {\it all} of  the near-extremal cases of current interest.

In summary, we have  been discussing the effect
of thermal fluctuations on the entropy and stability of a  black hole.
More specifically, we have reviewed a prior
canonical/grand canonical treatment \cite{NEW} that 
is novel for (at least) two reasons. 
Firstly, our methodology emphasizes the distinction between
systems with a fixed charge  and those for
which the charge is dynamical. Secondly, our
formalism  directly incorporates the black hole area spectrum,
 which should be, as we have argued \cite{NEW},
 regarded as an essential ingredient in any  calculation of this nature.  
Following this approach, we where able to explore the thermodynamic behavior
of black holes when  near certain critical points. For instance,
 it was shown that, due to thermal 
fluctuations, instability sets in  well before the system reaches
either one of
the {\it phase-transition } points; that is,  the 
Hawking-Page phase transition  \cite{haw4}
or  the  transition point of Reissner-Nordstrom black holes 
\cite{dav}. More precisely, in  both of these cases,
 instability will occur at the order 
of $A^{2/3}$ away from  the critical value of $A$. On the other hand, the
 thermal 
fluctuations are completely suppressed for a system close to extremality. 
Therefore, in a near-extremal regime, instability ensues 
at only a  few Planck scales below the  extremal point.

\section{Acknowledgments}

The authors graciously thank V.P. Frolov for
helpful conversations. GG is also grateful for
the Killam Trust for its financial support.

\end{document}